\documentclass[12pt]{article}
\usepackage{epsfig}
\usepackage{epsfig}
\begin{document}
\def\tr{{\rm tr}\, }
\def\Tr{{\rm Tr}\, }
\def\hTr{\hat{\rm T}{\rm r}\, }
\def\be{\begin{eqnarray}}
\def\ee{\end{eqnarray}}
\def\ctt{\chi_{\tau\tau}}
\def\cta{\chi_{\tau a}}
\def\ctb{\chi_{\tau b}}
\def\cab{\chi_{ab}}
\def\cba{\chi_{ba}}
\def\ptt{\phi_{\tau\tau}}
\def\pta{\phi_{\tau a}}
\def\ptb{\phi_{\tau b}}
\def\>{\rangle}
\def\<{\langle}
\def\d{\hbox{d}}
\def\pab{\phi_{ab}}
\def\lb{\label}
\def\appendix{{\newpage\section*{Appendix}}\let\appendix\section%
        {\setcounter{section}{0}
        \gdef\thesection{\Alph{section}}}\section}
\renewcommand{\figurename}{Fig.}
\renewcommand\theequation{\thesection.\arabic{equation}}
\hfill{\tt IUB-TH-058}\\\mbox{}
\vskip0.3truecm
\begin{center}
\vskip 2truecm {\Large\bf Holographic Description of Gravitational Anomalies}
\vskip 1.5truecm
{\large\bf
Sergey N.~Solodukhin\footnote{
{\tt s.solodukhin@iu-bremen.de}}
}\\
\vskip 0.6truecm
\it{School of Engineering and Science, \\
International University Bremen, \\
P.O. Box 750561,
Bremen 28759,
Germany}
\end{center}
\vskip 1cm
\begin{abstract}
\noindent
The holographic duality can be extended to include the quantum theories with the broken coordinate invariance
leading to the appearance of the gravitational anomalies. On the gravity side
one adds the gravitational Chern-Simons term to the bulk action which is gauge invariant  only up to the boundary terms.
We analyze in detail how the gravitational anomalies originate from the modified Einstein equations in the bulk.
As a side observation, we find that the gravitational Chern-Simons functional has the interesting conformal
properties. It is invariant under the conformal transformations. Moreover, its metric variation
produces a conformal tensor that is a generalization of the Cotton tensor to dimension $d+1=4k-1,~k\in Z$.
We calculate the modification of the holographic stress-energy tensor which is due to the Chern-Simons term
and use the bulk Einstein equations to find its divergence and thus reproduce the gravitational
anomaly. The explicit calculation of the anomaly is carried out in dimensions $d=2$ and $d=6$.
The result of the holographic calculation is compared with that of the descent method and an agreement is found.
The gravitational Chern-Simons term originates by the Kaluza-Klein mechanism from a one-loop
modification of  M-theory action. This modification is discussed in the context of the gravitational
anomaly in the six-dimensional $(2,0)$ theory. The agreement with the earlier conjectured anomaly is found.

\end{abstract}
\vskip 1cm
\newpage
\section{Introduction}
\setcounter{equation}0
Dualities play an important role in the theoretical concepts of modern physics.
In particular, they help to understand the behavior of certain systems at the strong
coupling by relating it to  the behavior in a weak coupling regime.
The AdS/CFT correspondence \cite{Maldacena:1997re}, \cite{Gubser:1998bc},
\cite{Witten:1998qj} (for review see \cite{Aharony:1999ti} and more recent \cite{deBoer:2004yu})
is the duality of this sort. Quite remarkably, it relates not only the different regimes but also
the apparently different theories. On one side of the duality one has superstring theory or M-theory, semiclassically
described by 11-dimensional supergravity, on the product of $AdS_{d+1}$ and a compact manifold.
On the other side it is the large N quantum strongly interacting conformal
theory living on the conformal boundary of  the Anti-de Sitter space. The duality works both ways.
It can be used to understand the strongly coupled quantum system in terms of the semiclassical gravitational
physics in the bulk. On the other hand, the wisdom gained in the long-time study of the quantum non-gravitational models
can be directed to solving the long-standing puzzles of the semiclassically quantized gravity. Among such puzzles
one finds the problem of  the black hole entropy and the unitarity problem.

\medskip

The duality in question has the interesting geometric aspects. As is known since the earlier works \cite{Brown:1986nw} and
\cite{FG}, there is conformal structure associated with  infinity of anti-de Sitter space.
Namely, one finds that the asymptotic symmetries which preserve the AdS structure also generate the conformal transformations
on the boundary at infinity.  On the other hand, the boundary metric serves as the Dirichlet data
for the boundary value problem associated with the bulk Einstein equations. Solution to this problem is a bulk metric
determined by the boundary data. This is one of the reasons why this duality is associated with
holography \cite{'tHooft:1993gx}, \cite{Susskind:1994vu}.
The latter states, quite generally,
that the fundamental degrees of freedom are that of the boundary
and predicts the possibility to project the bulk physics to the boundary.
In  Maldacena's picture the semiclassical gravitational action in the bulk becomes the
quantum generating functional for the theory on the boundary. In particular, its variation with respect to the
boundary metric (considered as a source for the dual stress-energy tensor) produces the n-point correlation functions
of the stress-energy tensor in the boundary theory.  The one-point function determines, for instance, the conformal
anomaly in the boundary theory. Thus, at least in principle, the classical geometry of  an asymptotically AdS  space provides us with
a complete solution to the quantum dual theory.
The bulk action has infra-red divergences
since it involves the integration over an infinite volume. These are the UV divergences on the boundary theory side.
Thus, for this procedure to work
the action should be properly regularized by adding the suitable counterterms. These and other questions were
actively studied in the literature, see
\cite{Henningson:1998gx}-\cite{Karch:2005ms}. The mathematical side
of the story was reviewed in \cite{Anderson:2004yi}.

\medskip

This line of research turned recently to a new interesting direction related to the possibility to understand
holographically the gravitational anomalies that may  arise in the dual theory \cite{Kraus}, \cite{SS}.
Indeed, the dual theory is generically chiral. The quantization of such a theory may break the coordinate invariance
and lead to the anomalies. These anomalies are well studied \cite{Alvarez-Gaume:1983ig}, \cite{Bardeen:1984pm}
and are known to appear in dimension $d=4k-2,\ k\in Z$. In two dimensions they arise in a theory in which the left and right central
charges are not equal. In six dimensions the gravitational anomaly arises, in particular, in the (2,0) theory.
In the weak coupling regime this theory is described by a certain tensor multiplet
theory while in the other regime the strongly interacting $(2,0)$ theory
describes $N$ coincident M5 branes.  Holographically, the gravitational anomaly
originates from the gravitational Chern-Simons term\footnote{By the gravitational CS term we mean the term for the Lorentz group
$SO(d+1)$ defined with respect to the spin connection which is on the other hand is completely defined in terms of the vielbein.
This is different from the Chern-Simons term for the AdS group $SO(d+1,1)$ considered in the context of AdS/CFT correspondence
in    \cite{BT}. The CS term in this case is polynomial in curvature and does not lead to the appearance of the
gravitational anomalies in the boundary theory.} which can be added to the gravitational bulk action.
This term is not gauge invariant, the non-invariance resides on the boundary.  This is origin  for the anomaly
in the boundary theory.
In fact, this mechanism  is similar to the known \cite{Witten:1998qj} holographic origin of the gauge field anomaly which relates it to
the appearance of the gauge field Chern-Simons action in the bulk. In this case the Chern-Simons term
is related by supersymmetry to the Einstein-Hilbert action and, thus, appears in the leading order in $N$.
On the other hand, the gravitational Chern-Simons term may originate by the Kaluza-Klein mechanism from a one-loop
modification of M-theory action \cite{Kraus:2005vz}.
Thus, the gravitational anomaly appears in the subleading order in $N$. It should be said that the quantum anomalies
are important, and sometimes the only one available, source of information about the strongly coupled theory.
That is why they  should be paid our special attention.

\medskip

In this paper we give an exhaustive analysis of the holographic gravitational anomaly.
Since there is no literature on the gravitational Chern-Simons terms beyond 3 dimensions
we start with a detail study of their general properties. In particular, we observe that
these are conformally invariant functionals. The field equations that follow from the
Chern-Simons action are, thus, traceless. This is not all, however, to the conformal properties.
The metric variation of the Chern-Simons term results in a conformal tensor.
This means that under the conformal transformations it rescales by a scalar factor.
The conformal  tensors that we have found exist in any dimension $d+1=4k-1, \ k\in Z$ and are different from the known Weyl tensors.
Since conformal tensors play an important and special role both in physics and mathematics
it would be interesting to see
if the tensors we have discovered have  their place in the available list
of conformal tensors.

\medskip

The modified Einstein equations in the bulk  are  subject to the Dirichlet problem.
We fix the boundary metric and solve the equations by doing the Fefferman-Graham expansion
for  the bulk metric. The full analysis of the problem is rather complicated. However,
in order to gain information about the divergence of the dual stress tensor we have to look
at a certain (dependent on the boundary dimension $d$) order in the expansion of   $(r,i)$ component of the Einstein equations.
This way we calculate the gravitational anomaly in $d=2$ and $d=6$. The holographic stress-energy tensor
is defined conventionally as a variation of the gravitational action with respect to the boundary metric.
We carry out the calculation of the stress-energy tensor that is due to the presence of the Chern-Simons term in the bulk action
and find a general expression for the tensor which is valid in any dimension $d$.
We then compare the holographic anomalies with
what one obtains in the standard descent method and find an agreement. Finally, we analyze the gravitational anomaly
in six dimensions as arising holographically from a one-loop modification of the gravitational action.
We compare it with the conjectured anomaly for the $(2,0)$ theory and find a complete agreement.
Before turning to the analysis let us emphasize that throughout the paper we consider  space-time of Euclidean signature
and use  the standard (see \cite{HE} for instance) conventions for the definition of the curvature.

\section{Brief review of gravitational anomalies}
\setcounter{equation}0
In this section our main source is the original paper \cite{Alvarez-Gaume:1983ig}
and the second volume of the book \cite{Green:1987mn}. A more recent review is \cite{Harvey:2005it}.
In the parity-preserving case one can always employ the Pauli-Villars regularization of loop diagram
which preserves the gauge invariance. The violation of gauge invariance occurs for fields whose
 gauge  couplings violate parity. This is the case for the general coordinate invariance for fields
which are in a complex (or pseudo-real) representation of the Lorentz group that violates parity.
In Euclidean signature, the complex representations of the Lorentz group $SO(d)$ of $d$-dimensional Minkowski space
occurs if dimension $d=4k-2$.

As other gauge anomalies, the gravitational anomaly has a topological origin and is related to certain topological invariants
of the tangent bundle in dimension $d+2=4k$. These invariants are polynomial in the Riemann curvature $R^a_{\ b}=
{1\over 2}R^a_{\ b \mu\nu}dx^\mu \wedge dx^\nu$
two-form. We remind that $R^a_{\ b}=d\omega^{a}_{\ b}+\omega^a_{\ c}\wedge \omega^c_{\ b}$, where
$\omega^a_{\ b}=\omega^a_{\ b,\mu}dx^\mu$ is the spin connection one-form,
and, with respect to indices $a$ and $b$,
is antisymmetric $d\times d$ matrix. An important property  is that the trace of any odd number of matrices $R^{a}_{\ b}$ vanishes,
$$
\Tr (R^{2k-1})=0~~.
$$
Thus only the even powers of $R$ can be used to construct the invariants. Since the latter should be further integrated over
a manifold $M$ of dimension $D$, only if  $D=4n$ these invariants are non-trivial. The gravitational anomaly   in dimension $d$
is obtained by the descent mechanism from the invariants in two dimensions higher, $D=d+2$.
This is yet another reason why the gravitational anomaly is expected to appear in the dimension $d=4k-2$.

There are certain combinations of  invariants constructed from $R$ which are sort of primary and are called
the Pontryagin classes. Notice, that
if $d$ is even, by an orthogonal transformation such an antisymmetric $d\times d$ matrix can be brought to a skew
diagonal form in terms
of its eigen-values $x_i~,i=1..{d\over 2}$. The characteristic Pontryagin class $p_i(M)$ is defined by
\be
&&\det (1-{1\over 2\pi }R)=\sum_{k=0}^\infty {p_k\over (2\pi)^{2k}} \nonumber \\
&&p_0(M)=1 \nonumber \\
&&p_1(M)\equiv \sum_i x^2_i=-{1\over 2}\Tr R^2 \nonumber \\
&&p_2(M)\equiv \sum_{i<j}x^2_ix^2_j=-{1\over 4}\Tr R^4+{1\over 8}(\Tr R^2)^2 \nonumber \\
&&p_3(M)\equiv \sum_{i<j<k}x^2_ix^2_jx^2_k=-{1\over 6}\Tr R^6+{1\over 8} \Tr R^2~\Tr R^4-{1\over 48}(\Tr R^2)^3~~.
\lb{1}
\ee
In the descent mechanism, just mentioned, one substitutes $R_{ab}$ in (\ref{1})  by $R'_{ab}=R_{ab}+\nabla_a\xi_b-\nabla_b \xi_a$
and then expands everything to the first order in $\xi_a$. Being integrated over a d-dimensional manifold
the result takes a general form
$$
\int d^dx \xi^\mu X_\mu~~,
$$
where $X_\mu$ is constructed via the Riemann tensor and its first derivative,
thus leading to  anomaly in the non-conservation of the stress-energy tensor
\be
\nabla_\alpha T^\alpha_{\ \mu}=X_\mu~~.
\lb{2}
\ee
The concrete form of $X_\mu$ depends on the dimension $d=4k-2$ and  the type of the field.
The anomaly for spin 1/2 particle is determined by  applying this mechanism  to the Dirac genus\footnote{In order to remove a common
factor  one usually defines $I_{1/2}=-i(2\pi)^{-D/2}\hat{I}_{1/2}$ with similar definitions of
$\hat{I}_{A}$.}
$$
\hat{I}_{1/2}=\prod_i{x_i/2\over \sinh(x_i/2)}~~.
$$
Its expansion in terms of the eigenvalues $x_i$ gives
\be
\hat{I}_{1/2}=1-{1\over 24}p_1+{1\over 5760}(7p_1^2-4p_2)+{1\over 2^6 15120}(-16p_3+44p_1p_2-31p_1^3)+..
\lb{3}
\ee
The anomaly for an antisymmetric self-dual tensor is described by
\be
\hat{I}_A=-{1\over 8}\prod_i{x_i\over \tanh x_i}
\lb{4}
\ee
which has expansion
\be
\hat{I}_A=-{1\over 8}-{p_1\over 24}+{1\over 5760}(16p_1^2-112p_2)+{1\over 967680}(-7936 p_3+1664p_1p_2-256p_1^3)+..~~.
\lb{5}
\ee
Another field which may contribute to the gravitational anomaly is gravitino.
Its anomaly is determined by the corresponding invariant polynomial. Since the theory on the boundary of AdS
is not supposed to contain gravity we skip the discussion of the anomaly due to gravitino.

In addition to the pure gravitational anomalies there may be the mixed anomalies that are due to
loop diagrams that contain both external gravitons and gauge fields. Thus, the chiral field should carry
the Yang-Mills charge. The only massless chiral field of this type is Weyl
spinor. The mixed anomaly then is determined by invariant polynomials involving both the curvature $R$  two-form
and the field strength $F=dA+A\wedge A$ of the Yang-Mills field. For gauge field in real representation of the gauge group we have that
$\tr F^{2k+1}=0$ and the relevant polynomial is
\be
\hat{I}_{1/2}(F,R)=\tr (\cos{F})\hat{I}_{1/2}(R)~~,
\lb{6}
\ee
where $\hat{I}_{1/2}(R)$ was introduced above.
It has the following expansion
\be
 \hat{I}_{1/2}(F,R)=n+[c_2-{n\over 24}p_1]+[-{1\over 6}(c_4+{1\over 2}c_2^2)+{n\over 5760}(7p_1^2-4p_2)-{p_1\over 24}c_2]+..~~,
\lb{7}
\ee
where $n=\tr 1$ is dimension of the representation of the gauge group and $c_j(F)$ is the Chern class defined as
$\det (1+iF/2\pi)=\sum_ji^jc_j(F)/(2\pi)^j$. In terms of the field strength we have that
$c_0(F)=1$, $c_2(F)=-{1\over 2}\tr F^2$, $c_4(F)={1\over 8}(\tr F^2)^2-{1\over 4}\tr F^4$.

\section{Gravitational Chern-Simons terms}
\setcounter{equation}0
The gravitational Chern-Simons terms $\Omega_{2n+1}$ are defined
as\footnote{From now on we will suppress symbol $\wedge$ for
the wedge product of several differential forms.}
\be
d\Omega_{2n+1}=\Tr R^{n+1}
\lb{2.1}
\ee
and are certain polynomials of the spin connection $\omega^a_{\ b}$ and
its exterior derivative $d\omega^a_{\ b}$ (or, equivalently, of curvature $R^a_{\ b}$).
A closed form for arbitrary $n$ is
\be
\Omega_{2n+1}=(n+1) \int_0^1\ dt \ t^{n} \Tr (\omega (d\omega+t\omega^2)^{n})~~.
\lb{2.2}
\ee
Both the spin connection $\omega^a_{\ b}$ and the curvature $R^a_{\ b}$ take values in the algebra of
the Lorentz group so that the other name for $\Omega_{2n+1}$ is the Lorentz Chern-Simons term. Thus, both
$\omega$ and $R$ are antisymmetric in the Lorentz indices.
Variation of the term (\ref{2.2}) under a small change of the spin connection is
\be
\delta \Omega_{2n+1}=(n+1)\Tr(\delta\omega R^{n})+d(...)~~,
\lb{2.3}
\ee
where $d(...)$ stands for a term which is exact  form.
As was  discussed in section 2, $R$ is antisymmetric matrix so that the trace of the product of odd
number of $R$ gives zero. Similarly, we have that $\Tr (\delta \omega R^{2k})=0$ that can be
shown by taking the transposition of this expression. Thus, for even $n$  the right hand side of
both (\ref{2.1}) and (\ref{2.3}) is vanishing (in the case of (\ref{2.3}) it is up to an exact form).
So that,  action $\int \Omega_{2n+1}$ does not produce  any non-trivial field equations if $n$ is even.
The case of odd $n=2k-1$ will be further considered. The Chern-Simons action
\be
W_{\tt CS}^{}=a_n\int_{M^{2n+1}} \Omega_{2n+1}~~,~~a_n={2^n\over n+1}~~,
\lb{2.4}
\ee
where $n=2k-1$ with integer $k$, describes the non-trivial dynamics for the gravitational field.

The spin connection is not independent variable. It is determined by equation
\be
de^a+\omega^a_{\ b}\wedge e^b=0~~,
\lb{2.5}
\ee
where $e^a=h^a_\mu dx^\mu$ is  the vielbein, a "square root" of metric, $G_{\mu\nu}=h^a_\mu h^b_\nu \delta_{ab}$.
 The components of the vielbein can be used to project the
local Lorentz indices to the coordinate indices and vice versa.
 Useful formula for the calculation of the components of the spin connection in
terms of the vielbein is
\be
&&\omega_{ab,\mu}={1\over 2}(C_{a\nu\mu}h^\nu_b+C_{b\mu\nu}h^\nu_a-C_{d\alpha\beta}h^\alpha_ah^\beta_bh^d_\mu)~~, \nonumber \\
&&C^a_{\mu\nu}\equiv \partial_\mu h^a_\nu-\partial_\nu h^a_\mu~~.
\lb{2.6}
\ee
The Riemann curvature satisfies the two types of identities
\be
R^a_{\ [\mu,\alpha\beta]}=0~~&\Longleftrightarrow& ~~R^a_{\ b}\wedge e^b=0~~(1) \nonumber \\
\nabla_{[\alpha}R^{\mu\nu}_{\ \ \beta\gamma]}=0~~&\Longleftrightarrow& ~~~\nabla R^a_{\ b}=0 \ \ \ \  (2)
\lb{Bianchi}
\ee
which will be useful in our analysis.

\bigskip

\noindent{\bf Conformal invariance.} In this section we would like to find a general form for the field equations which follow from the Chern-Simons
action (\ref{2.4}) when we vary the vielbein. This will be done in a moment. We pause here to show that
the gravitational Chern-Simons is actually a conformal invariant so that the field equations that follow from (\ref{2.4}) should be traceless.
It immediately follows from (\ref{2.6}) that under the rescaling of the vielbein, $h^a_\mu\rightarrow e^\sigma h^a_\mu$,
the components of spin connection change as
\be
\omega_{ab,\mu}\rightarrow \omega_{ab,\mu}+\partial_b \sigma h_{a\mu}-\partial_a\sigma h_{b\mu}~~,
\lb{2.7}
\ee
where we define $\partial_a\equiv h_a^\mu\partial_\mu$. The conformal variation of the bulk part of
the Chern-Simons term (\ref{2.3})
vanishes due to the Bianchi (1) identity. The action (\ref{2.4}) is thus conformally invariant provided that the conformal parameter
$\sigma$ vanishes on the boundary of $M^{4k-1}$. This is an interesting feature common to  the gravitational Chern-Simons terms
in all dimensions.

\bigskip

\noindent{\bf Field equations.} Now we are in a position to find an explicit form for the field equations which follow from
the Chern-Simons action (\ref{2.4}) when we vary the vielbein $h^a_\mu$. We first rewrite the integrated variation formula
(\ref{2.3}) in components and  neglect possible boundary terms\footnote{We use
  that $dx^{\sigma_1}\wedge ..\wedge
  dx^{\sigma_{2n+1}}=\epsilon^{\sigma_1..\sigma_{2n+1}}h d^{2n+1}x$, $h=\det
  h^a_\mu$ and $\epsilon^{\sigma_1\sigma_2..}=h^{\sigma_1}_{a_1}h^{\sigma_2}_{a_2}..\epsilon^{a_1a_2..}$.}
\be
\delta W^{}_{\tt CS}=\int d^{2n+1}x \ h \ \epsilon^{\sigma_1\sigma_2...\sigma_{2n}\mu}\ R^a_{\ a_1 \sigma_1 \sigma_2}R^{a_1}_{\ a_2 \sigma_3\sigma_4}..
\ R^{a_{2n-2}}_{\ b \ \sigma_{2n-1}\sigma_{2n}} \delta \omega^b_{\ a, \mu}~~,
\lb{2.8}
\ee
where $h=\det h^a_\mu$.
A variation of the spin connection (\ref{2.6}) under an infinitesimal change of the vielbein
\be
\delta \omega^a_{\ b,\mu}=\delta \Gamma^\alpha_{\mu\nu}h^a_\alpha h_b^\nu-h_b^\nu \nabla_\mu \delta h^a_\nu
\lb{2.9}
\ee
is a
combination of a part due to the variation of the vielbein alone and of
another part which is due to the variation of
the vielbein inside the metric. The latter comes from the
variation of the Christoffel symbol
\be
\delta \Gamma^\alpha_{\mu\nu}={1\over 2}[-\nabla^\alpha\delta g_{\mu\nu}+\nabla_\mu \delta g^\alpha_{\ \nu}+\nabla_\nu
\delta g^\alpha_{\ \mu}]~~,~~\delta g^\alpha_{\ \mu}\equiv g^{\alpha\nu}\delta g_{\mu\nu}~~.
\lb{Gamma}
\ee
Substituting (\ref{2.9}) into (\ref{2.8}) we notice that the part due to the variation of the vielbein vanishes after
integrating by parts and using the Bianchi (2) identities. The only non-trivial variation thus comes from that of
the metric. This variation of the Chern-Simons term can be shown to vanish (provided that both types of the Bianchi identities
are used)
identically if $n$ is even. This is of course consistent with the  arguments given earlier in this section. If $n$ is odd
the variation is non-trivial
\be
\delta W^{}_{\tt CS}=-2\int_{M^{2n+1}}d^{2n+1}x \ h \ \delta g_{\mu\nu} \ C^{\mu\nu}
\lb{2.10}
\ee
with a tensor $C^{\mu\nu}_{}$ defined as
\be
&&C^{\mu\nu}_{}=\nabla_\alpha S^{(\mu\nu)\alpha}~~,\lb{2.11} \\
&&S^{\mu\nu\alpha}=-{1\over 2} \epsilon^{\sigma_1\sigma_2...\sigma_{2n}\mu}\ R^{\nu }_{\ a_1 \sigma_1 \sigma_2}R^{a_1}_{\ a_2 \sigma_3\sigma_4}...
\ R^{a_{2n-2}\alpha}_{\ \ \sigma_{2n-1}\sigma_{2n}}~~, \nonumber 
\ee
where symmetrization  is defined as $B^{(\mu\nu )}={1\over 2}(B^{\mu\nu}+B^{\nu\mu})$.
The tensor $S^{\mu\nu\alpha}$ is antisymmetric in last two indices. It is vanishing when the trace over any pair of indices
is taken  and is covariantly conserved,
\be
S^{\mu\nu\alpha}=-S^{\mu\alpha\nu}~,~~S^{\alpha\nu}_{\ \  \ \alpha}=0~,~~\nabla_\mu S^{\mu\nu\alpha}=0~~.
\lb{SS}
\ee
In this respect it resembles a tensor of
spin. We however do not pursue this analogy in the present paper.
By virtue of the Bianchi identities (\ref{Bianchi}) the tensor $C^{\mu\nu}$  is  traceless and covariantly conserved.

\bigskip

\noindent {\bf Dimension d+1=3 ($n=1$).}
The three-dimensional General Relativity with  the
gravitational Chern-Simons term added is known as a topologically massive gravity and was first considered
in \cite{Deser1} and \cite{Deser2}.
In three dimensions   we have that
\be
S^{\mu\nu\alpha}=-{1\over 2}\epsilon^{\sigma_1\sigma_2\mu}R^{\nu\alpha}_{\ \  \sigma_1\sigma_2}~~.
\lb{S2}
\ee
This can be further brought to another form using the fact that the Riemann
tensor in three dimensions is expressed in terms of the Ricci tensor and the
Ricci scalar as follows
$$
R^{\nu\alpha}_{\ \ \sigma \rho}=\delta^\nu_{\sigma} P^\alpha_\rho
+\delta^\alpha_{\rho} P^\nu_\sigma -\delta^\nu_{\rho}
P^\alpha_\sigma -\delta^\alpha_{\sigma} P^\nu_\rho~~,
$$
where $P^\alpha_\beta=R^\alpha_\beta-{1\over 4}\delta^\alpha_\beta R$.
By means of this relation we find that
\be
S^{\mu\nu\alpha}=\epsilon^{\sigma\nu\mu}P^\alpha_\sigma+\epsilon^{\mu\alpha\sigma}
P^\nu_\sigma~~.
\lb{S21}
\ee
The first term in the above expression is antisymmetric in $\mu$ and $\nu$ so
it drops out in the symmetrization (\ref{2.11}). The second term, on the
other hand, is symmetric in indices $\mu$ and $\nu$ that can be shown by
contracting this term with $\epsilon_{\mu\nu\rho}$ and demonstrating that this gives
zero provided the Bianchi identities are employed once again. We finally have that
\be
C^{\mu\nu}=\nabla_\alpha S^{(\mu\nu)\alpha}=\epsilon^{\mu\alpha\sigma}\nabla_\alpha (R^\nu_\sigma
-{1\over 4}\delta^\mu_\sigma R)~~.
\lb{2.11*}
\ee
In three dimensions the tensor $C^{\mu\nu}$ is known as the Cotton tensor. It plays an important role
since it is the only conformal tensor available in three dimensions.
Expression (\ref{2.11}) gives a generalization\footnote{There have been
earlier suggested some generalizations \cite{Garcia:2003bw} of the Cotton tensor to higher dimensions.
These are however linear in the Riemann curvature and thus differ from (\ref{2.11}).}
of the Cotton tensor to higher dimensions ($n>1$).
The  higher dimensional generalizations
give the conformal tensors as well as we know discuss.

\bigskip

\noindent{\bf Conformal property of $C^{\mu\nu}$.}
The tensor $C^{\mu\nu}$ defined in (\ref{2.11}) is a conformal
tensor of weight $-(d+3)$. This property makes it similar to Weyl tensor.  In order to obtain the
transformation law for the tensor $C^{\mu\nu}$  in dimension $d+1$
($d=2n=4k-2, k\in Z$) we first
note that under an infinitesimal conformal transformation
$\delta_\sigma h^a_\mu=\delta\sigma h^a_\mu$ the tensor $S^{(\mu\nu)\alpha}$ transforms as follows
\be
\delta_\sigma S^{(\mu\nu)\alpha}=-(d+3)\delta\sigma \ S^{(\mu\nu)\alpha}
+\epsilon^{\sigma_1..\sigma_{d-1}\alpha (\mu}
R^{\nu )}_{\ a_1 \sigma_1\sigma_2}..R^{a_{n-2}}_{\ a_{n-1}\sigma_{d-3}\sigma_{d-2}}
\nabla_{\sigma_{d-1}}\nabla^{a_{n-1}}\delta\sigma ~.
\lb{dS}
\ee
{}From this it is straightforward to derive that
\be
\nabla_\alpha \{ \delta S^{(\mu\nu)\alpha}\}=-(d+3)\delta\sigma \ S^{(\mu\nu)\alpha}-
(d+2)S^{(\mu\nu)\alpha}\ \partial_\alpha  \delta\sigma~~.
\lb{ddS}
\ee
Combining this with an obvious property
\be
\delta(\nabla_\alpha)S^{(\mu\nu)\alpha}=(d+2)S^{(\mu\nu)\alpha}\ \partial_\alpha\delta\sigma
\lb{SSd}
\ee
we find that tensor $C^{\mu\nu}=\nabla_\alpha S^{(\mu\nu)\alpha}$ transforms as
\be
\delta C^{\mu\nu}=-(d+3)\delta \sigma \ C^{\mu\nu}
\lb{dC}
\ee
under the conformal transformations. As is well known (see, for instance, Proposition 2.1 in \cite{GW})
the transformation law (\ref{dC})  under the infinitesimal conformal transformations implies that
the tensor $C_{\mu\nu}$ is   conformal and changes properly under the finite conformal transformations.
On the other hand, this property follows directly from the
fact that $C^{\mu\nu}$ is obtained as a metric variation of a conformally invariant
functional (a nice discussion of this general fact can be found, for instance,  in \cite{Skenderis:2000in}).
The tensor $C^{\mu\nu}$, thus,
vanishes for any metric conformal to the maximally symmetric, constant
curvature, metric $g^{\tt cc}_{\mu\nu}$.
Indeed, the Riemann tensor $R^{\alpha\beta}_{\ \ \mu\nu}={R\over d(d+1)}(\delta^\alpha_\mu \delta^\beta_\nu-\delta^\alpha_\nu\delta^\beta_\mu)$
for a maximally symmetric metric so that the tensor $S^{\mu\nu\alpha}$, and hence $C^{\mu\nu}$,
vanishes identically in this case.

Thus, the  tensors $C_{}^{\mu\nu}$ (\ref{2.11}) share same properties in all
dimensions $4k-1,~k\in Z$:
they are  traceless, covariantly conserved and conformal. Conformal tensors
traditionally play a special role in differential geometry
and their complete classification is a long-standing problem. We are, however, not aware of any
earlier appearance of tensors (\ref{2.11}) in the mathematics or physics
literature.

This tensors, actually, differ from all known conformal tensors in an
interesting way.  Consider metric $g_{\mu\nu}=g^{\tt cc}_{\mu\nu}+\eta_{\mu\nu}$ that is a small deformation  of a
constant curvature maximally symmetric metric  $g^{\tt cc}_{\mu\nu}$. Usually, a conformal
tensor $T$ for such a deformation takes the form (skipping the indices) $T={\cal
  D} \eta$ with ${\cal D}$ being some invariant differential operator. Such
operators can be  classified that allows to classify all conformal invariants
that are represented in such a form for a small deformation of the  maximally
symmetric metric. The corresponding classification theorem is due to Graham
and Hirachi \cite{Graham-Hirachi}. In particular, it says that in odd
dimensions  there is only Weyl tensor. Interestingly,
tensor $C^{\mu\nu}$ (\ref{2.11}) does not fit in the conditions of this
theorem\footnote{I thank Robin Graham for discussions on this point.}.  It is polynomial in the small deformation of the
maximally symmetric
metric: $C[g^{\tt cc}+\eta]\sim \eta^{2k-1}$ in dimension $4k-1$, $k>1$.
This can be easily seen already for tensor $S^{\mu\nu}$: due to the Bianchi
identities the linear term and all terms of order $\eta^{2l-1},\ l<k$ vanish
identically. In dimension 7 one can find an explicit form for the leading term.
It is more convenient to write it for the tensor $S^{\mu\nu\alpha}$,
\be
&&S^{(\mu\nu)\alpha}=-{1\over 2}\epsilon^{\sigma_1 ..\sigma_6
  (\mu} D^{\nu)}_{\ \sigma_1 a_1\sigma_2}D^{a_1}_{\
  \sigma_3 a_2\sigma_4}D^{a_2\ \alpha}_{\ \sigma_5 \ \sigma_6}\nonumber \\
&&+{R\over 2d(d+1)}
\epsilon^{\sigma_1 ..\sigma_5\alpha (\mu}D^{\nu)}_{\ \sigma_1 a_1\sigma_2}D^{a_1}_{\
  \sigma_3 a_2\sigma_4}\eta^{a_2}_{\ \sigma_5}~~,
\lb{Sh3}
\ee
where we have introduced notation
$$
D^{\alpha}_{\ \sigma_1 a\sigma_2}=\nabla_{\sigma_1}(\nabla_a\eta^\alpha_{\
  \sigma_2}-\nabla^\alpha\eta_{a\sigma_2})~~,~~\eta^\alpha_{\ \sigma}=g^{\alpha\beta}_{\tt cc}\eta_{\beta\sigma}~~.
$$

\bigskip

\noindent{\bf Reducible Chern-Simons terms.} So far we  considered the irreducible form of the Chern-Simons terms.
There can be, however, forms which reduce to the product of several such terms. An example is
\be
W^{(k,p)}_{\tt CS}=(n+1)a_n\ \int_{M^{2n+1}}\Omega_{2k+1}d\Omega_{2p+1}~~,~~n=k+p+1~~.
\lb{2.12}
\ee
A metric variation
\be
\delta W^{(k,p)}_{\tt CS}=-8\int_{M^{2n+1}} C^{\mu\nu}_{(k,p)}\delta
g_{\mu\nu}
\lb{Wkp}
\ee
of this action gives a tensor
\be
&&C^{\mu\nu}_{(k,p)}= -{1\over 8}\epsilon^{\sigma_1...\sigma_{2n}(\mu}\nabla_\alpha [(k+1)R^{\nu )}_{\ a_1 \sigma_1\sigma_2}..R^{a_{2k-2}\alpha}_{\ \ \sigma_{2k-1}\sigma_{2k}}
(R^{c_1}_{\ c_2 \sigma_{2k+1}\sigma_{2k+2}}..R^{c_{2p+1}}_{\ c_{2p+2}\sigma_{2n-1}\sigma_{2n}}) \nonumber \\
&&+(p+1)R^{\nu )}_{\ a_1 \sigma_1\sigma_2}..R^{a_{2p-2}\alpha}_{\ \ \sigma_{2p-1}\sigma_{2p}}
(R^{c_1}_{\ c_2 \sigma_{2p+1}\sigma_{2p+2}}..R^{c_{2k+1}}_{\ c_{2k+2}\sigma_{2n-1}\sigma_{2n}})]~~.
\lb{2.13}
\ee
It is traceless and covariantly conserved and is yet another possible generalization of the Cotton tensor to higher dimensions.
If one includes the Yang-Mills field into  consideration there may appear the mixed terms like
\be
W_{\tt mix}=\int_{M^{2n+1}}\Omega_{2p+1}\tr F^{k}~~,~~n=k+p~~.
\lb{2.14}
\ee
The  metric variation of this action is obvious.

\section{Holographic evaluation of  gravitational anomaly}
\setcounter{equation}0
According to  the holographic conjecture the (d+1)-dimensional gravitational theory (referred as the bulk
theory) is  equivalent to a d-dimensional conformal field (boundary) theory.
The boundary in question is the boundary of an asymptotically AdS space-time that is a solution to the
gravitational bulk theory. More generally, the duality is formulated for string theory (or M-theory)
on anti de-Sitter space, the (super)-gravity action is a low-energy approximation to this more fundamental theory.
The (super)-gravity action generically has the higher derivative modifications of the purely gravitational part of the action.
Here we consider the case when this modification is in the form of the gravitational Chern-Simons terms.
These terms may appear in particular due to the Kaluza-Klein reduction of the higher
curvature terms generically present
in  11-dimensional M-theory
action.

The gravitational theory in (d+1)-dimensional space-time is given by the action
\be
W_{\tt gr}=W_{\tt EH}-{\beta \over 32\pi G_N} W_{\tt CS}~~,
\lb{3.1}
\ee
which is sum of the Chern-Simons term (\ref{2.4}) and  the ordinary Einstein-Hilbert action (with a negative cosmological constant)
\be
W_{\tt EH}=-{1\over 16 \pi G_N}[\int_{M^{d+1}} (R[G]+d(d-1)/l^2)+\int_{\partial M^{d+1}}2K]~~,
\lb{3.2}
\ee
where $K$ is trace of the second fundamental form of boundary $\partial M$. $G_N$ is Newton's constant
in $d+1$ dimensions.
Parameter $l$ sets the AdS scale. We will use units $l=1$.
One can add to the action (\ref{3.1}) the reducible
forms of the gravitational Chern-Simons term existing in the dimension $d+1$. Note,
that the analytic continuation of the Chern-Simons action to Lorentzian signature
is somewhat subtle and involves the multiplication by $i$. So that if the coupling
$\beta$ is purely imaginary in Euclidean signature (as is reasonable from the
boundary point of view since the
gravitational anomaly comes from the imaginary part of the quantum action) it
becomes real in Lorentzian signature. The analytic continuation of the
topological terms  is
discussed in \cite{Bilal:2003es}.

The gravitational bulk equations obtained by varying the action (\ref{3.1}) with respect to the metric
takes the form
\be
R_{\mu\nu}-{1\over 2}G_{\mu\nu}R-{d(d-1)\over 2}G_{\mu\nu}+\beta C_{\mu\nu}=0~~,
\lb{*}
\ee
where all curvature tensors are determined with respect to the bulk metric $G_{\mu\nu}$. The tensor $C_{\mu\nu}$  is a result
of the variation of the  gravitational Chern-Simons term.
Although the Chern-Simons terms  are defined in terms of the Lorentz connection that is not gauge invariant object
the variation is presented in the covariant and gauge invariant form as we have shown in the previous section.
This is just a manifestation of the fact
that the "non-invariance" of the Chern-Simons term resides on the boundary  and does not appear in the bulk field equations.
By virtue of the Bianchi identities this quantity (both for the irreducible and reducible  Chern-Simons terms)
is  manifestly traceless and identically
covariantly conserved,
\be
C_{\mu\nu}G^{\mu\nu}=0~,~~\nabla_\mu C^{\mu}_{\ \nu}=0~~.
\ee
Due to  these properties we find that a solution to the equation (\ref{*}) is space-time with
constant Ricci scalar $R=-d(d+1)$. This is exactly what we had when the Chern-Simons term was not included in the action.
In that case moreover the Ricci tensor  was proportional to the metric,  $R_{\mu\nu}=-dG_{\mu\nu}$.
It is no more the case in the presence of the Chern-Simons term and we have
\be
R_{\mu\nu}=-dG_{\mu\nu}-\beta C_{\mu\nu}~~.
\lb{3.5}
\ee
This is that equation which we are going to solve.
We start with choosing the bulk metric in the form
\be
ds^2=G_{\mu\nu}dX^\mu dX^\nu=dr^2+g_{ij}(r,x)dx^idx^j
\lb{3.6}
\ee
that always can be done by using the  normal coordinates.
The quantity $g_{ij}(r,x)$ is the induced metric on the hypersurface of a constant value of the radial coordinate $r$.
The following expansion
\be
g(r,x)=e^{2r}[g_{(0)}+g_{(2)}e^{-2r}+..+g_{(d)}e^{-dr}+h_{(d)}~re^{-dr}+O(e^{-(d+1)r})]
\lb{3.7}
\ee
is  assumed
so that the metric (\ref{3.6}) describes an asymptotically anti-de Sitter space-time with $g_{(0)}$ being the metric on its
d-dimensional boundary. The non-vanishing term $h_{(d)}$ generically appears in the expansion if dimension $d$ is even.
In the mathematics literature this tensor is known as the obstruction tensor (see  \cite{Graham-Hirachi}, \cite{Gover}). It is traceless, covariantly conserved
and conformal  in any even dimension $d$.
By a general argument given in \cite{CMP} it is a multiple of the stress tensor derived from the
integrated holographic conformal anomaly. It follows immediately that this term vanishes identically when $d=2$
since the conformal anomaly then is a multiple of the Ricci scalar and, if integrated, gives a topological invariant
so that no non-trivial metric variation appears.

\bigskip

\noindent {\bf Holographic stress-energy tensor.}
The holographic (or dual) stress-energy tensor
is generally defined as a variation of the gravitational action with respect to the
metric $g^{(0)}_{ij}(x)$ on the boundary. The gravitational action is
considered on-shell, i.e. the bulk metric is supposed to solve the Einstein
equations subject to the Dirichlet boundary condition.
Since the boundary is at infinity the action should be properly defined.
A simple way to do it is to consider a sequence of boundaries at finite value of the radial
coordinate $r$ with induced boundary metric $g_{ij}(x,r)$ (for large $r$ we know that
$g_{ij}(x,r)=e^{2r}g_{(0)ij}(x)+..$). This way we get a regulated
gravitational action. However, this action is typically divergent when regulator $r$ is taken to infinity.
On the boundary theory side
these divergences have a natural interpretation as the UV divergences. Some {\it renormalization} is
typically needed. A rather natural way to {\it renormalize} the divergences is to add some local
boundary counter terms \cite{Henningson:1998gx}, \cite{Balasubramanian:1999re},
\cite{Kraus:1999di}, \cite{CMP} to the action.  These boundary terms do not
change the bulk field equations. They not just cancel the divergences but also
contribute to the finite part of the action and, in particular, to the
finite part of the  holographic stress-energy tensor. When the boundary dimension
$d$ is odd the exact form of the holographic stress tensor is known \cite{CMP}. It is
determined only by the coefficient $g^{(d)}_{ij}(x)$ in the Fefferman-Graham. 
When  dimension $d$ is even no
general form of the dual stress-energy tensor is known except in some particular cases:
$d=2$, $d=4$ and $d=6$ \cite{CMP}. The expression in terms of the extrinsic curvature
(instead of the metric) is, however, available \cite{Papadimitriou:2004ap}.

In the presence of the Chern-Simons term the holographic stress-energy tensor
is modified. Surprisingly, we can get a general form for this modification
rather explicitly. In order to see this let us remind the basic steps
in defining the holographic stress-energy tensor.
Let us introduce a small parameter $\varepsilon=e^{-2r}$ which determines the
location of the regularized boundary with the induced metric $g_{ij}(x,r(\varepsilon))$
which is the same quantity that appears in (\ref{3.6}).
The expectation value of the stress-energy tensor of the dual theory is then
given by
\be
<T_{ij}>={2\over \sqrt{\det g_{(0)}}} {\delta W_{\tt gr,ren}\over
      \delta g^{ij}_{(0)}(x)}=\lim_{\varepsilon\rightarrow 0}\left({1\over
        \varepsilon^{d/2-1}}T_{ij}[g]\right)~~,
\lb{ht1}
\ee
where
\be
T_{ij}[g]={2\over \sqrt{\det{g(x,r(\varepsilon))}}}{\delta W_{\tt gr,ren}\over
      \delta g^{ij}(x,r(\varepsilon))}
\lb{ht2}
\ee
is the stress tensor of the theory at finite $\varepsilon$. It contains two
contributions,
$$
T_{ij}[g]=T^{\tt reg}_{ij}+T^{\tt ct}_{ij}~~,
$$
where $T^{\tt reg}_{ij}=-{1\over 8\pi G_N}(K_{ij}-Kg_{ij})$ comes from the
regulated Einstein-Hilbert action and
$T^{\tt ct}_{ij}$ is the contribution of the boundary counterterms, their role
is to cancel the possible divergences in (\ref{ht1}) when
$\varepsilon$ is taken to zero.

We now want to apply this prescription and compute the
dual stress-energy tensor which is due to the gravitational Chern-Simons term
in the bulk. All we need
to do is take a variation  of the Chern-Simons action with respect to the
induced metric $g^{ij}(x,r(\varepsilon))$  on the regularized boundary and
calculate $T^{}_{ij}[g]$. Then
insert it into equation (\ref{ht1}) and take the limit of $\varepsilon$ to zero.
The equation (\ref{2.8}) accompanied with (\ref{2.9}) and (\ref{Gamma}) is a
good starting point for the first step. Notice, that in (\ref{2.8}) the variation with respect to
the boundary value of vielbein results in some term $T^a_i$ that,
however, vanishes after the symmetrization, $h_{a(j}T^a_{i)}$, needed  to define the metric
stress-energy tensor. Thus, the only contribution to the stress-energy tensor
comes from the metric variation in (\ref{Gamma}). We find that
\be
&&\delta W^{\tt reg}_{\tt CS}=-{\beta\over 32\pi G_N}\int d^dx
\sqrt{g}\ [S^{ijr}-S^{rji}-S^{irj}]\delta g_{ij} \nonumber \\
&&=-{\beta\over 16\pi G_N}\int
d^dx \sqrt{g} \ S^{(ij)r}\delta g_{ij}~~,
\lb{ht3}
\ee
where tensor $S^{\mu\nu\alpha}$ was introduced in (\ref{2.11}) and, in the
second line,  we have used
its symmetry properties (\ref{SS}). This gives us that
\be
T_{ij}[g]={\beta\over 8\pi G_N}S_{(ij)r}(g(x,r(\varepsilon)))~~.
\lb{ht4}
\ee
This should  then be expanded in the powers of $\varepsilon$ and  substituted in
(\ref{ht1}). The analysis shows that the leading divergences in the resultant
expression vanish either as a result of the symmetrization in indices $i$ and $j$  or
due to the Bianchi identities so that we are left with a finite
expression. This quite remarkable fact (in dimension $d=2$ this was observed
in \cite{Kraus}) means that the dual stress-energy tensor is finite with no need
to introduce any new counterterms.
The general form for the finite expression can be easily obtained for arbitrary $d$
by using the expansions
(\ref{R1}), (\ref{R2}) and (\ref{R3}) of Appendix B,
\be
T^{\tt CS}_{ij}=-{\beta\over 8\pi G_N}\epsilon^{k_1k_2..k_{d-1}}_{\ \ \ \ \ \
  \ \
    (i}R_{j)\ \ k_1k_2}^{\ n_1}R_{n_1\ k_3k_4}^{\ n_2}.. R_{n_{d-3}\ \
    k_{d-3} k_{d-2}}^{\ \ n_{d-2}}\ g^{(2)}_{n_{d-2}k_d}~~,
\lb{ht5}
\ee
where one uses the metric $g_{(0)ij}$ to compute the components of the Riemann tensor.
Thus, we have to know only the coefficients $g_{(0)ij}$ and $g_{(2)ij}$ in the
Fefferman-Graham expansion in order to determine the part in the dual stress-energy tensor
which is due to the Chern-Simons term.

\bigskip

\noindent {\bf  Solving the Einstein equations.} The expressions for the components of the bulk curvature are given in appendix A.
The strategy of solving the Einstein equations is to substitute the expansion (\ref{3.7}) into the modified
Einstein equations (\ref{3.5}) and expand both sides of the equations in powers of $e^{-r}$. Equating
coefficients at the same order on both sides one gets  the recurrent  relations between coefficients of the expansion
(\ref{3.7}) which allow one to determine
$g^{(n)}_{ij}(x)$ provided coefficients $g^{(k)}_{ij}(x),~k<n$ are already known. The only boundary data required for this procedure to
work is the value of the boundary metric $g_{ij}^{(0)}(x)$ and the value of the coefficient $g^{(d)}_{ij}(x)$ that is ultimately
related to the stress-energy tensor of the boundary CFT. Einstein equations impose constraints on  trace
and divergence of $g^{(d)}_{ij}$. The latter thus determines the conservation (or non-conservation) of the stress-energy  tensor.
The constraint on divergence of $g^{(d)}_{ij}$ appears in the $e^{-dr}$ order of the expansion of
 $(ri)$ component of the Einstein equations. It is thus suffice for our purposes to look only at this part of the
Einstein equations.
The expansion for the inverse metric and the Riemann tensor is given in appendix B.
Since we have to calculate the expansion of $(ri)$ component of the tensor $C^{\mu\nu}$ we
present below the expression for this component in terms of the tensor $S^{ijk}$,
\be
C^{ri}={1\over 2}\{\nabla_jS^{rij}+\nabla_jS^{irj}+\partial_r S^{rir}+{1\over 2}\Tr (g^{-1} g')S^{rir}+{1\over 2}(g^{-1}g')^i_jS^{rrj}
-{1\over 2}g'_{kn}S^{kin}\},
\lb{C}
\ee
where $\nabla_j$ is defined with respect to the induced metric $g_{ij}(r,x)$.
The further analysis depends on the value of dimension $d$.

\bigskip

\noindent {\it Dimension d=2.} The case of two-dimensional boundary was
considered in \cite{SS}  and the holographic tensor was found earlier in
\cite{Kraus}. Below we present some details of the analysis.
 In this case the first non-vanishing contribution to component $C^{ri}$ appears in
$e^{-4r}$ order. The components of tensor $S^{\mu\nu\alpha}$ are easy to calculate using (\ref{S2}) and the expansions
(\ref{R1}), (\ref{R2}), (\ref{R3}) of the Riemann tensor. In the leading order one has
\be
&&S^{rij}=(-{1\over 2}R\epsilon^{ij}+\epsilon^{ki}g_{(2)k}^j-\epsilon^{kj}g_{(2)k}^i)e^{-4r}+..  \\
&&S^{irj}=\epsilon^{ji} e^{-2r}+O(e^{-6r}) \nonumber \\
&&S^{rrj}=-\epsilon^{kn}\nabla_k g^j_{(2)n}e^{-4r}+.. \nonumber \\
&&S^{kin}=\epsilon^{kl}(\nabla^ig^n_{(2)l}-\nabla^ng^i_{(2)l})e^{-6r}+.. \nonumber
\lb{Sd2}
\ee
So that the leading term in the expansion of the component of the tensor $C^{\mu\nu}$ can be now calculated using
(\ref{C}),
\be
C_{ri}=g_{ij}C^{rj}=\{-{1\over 4}\epsilon_i^{\ j}\partial_j R+{1\over
  2}(\epsilon^{k}_{\ i}\nabla_jg_{(2)k}^j+
\epsilon^{kj}\nabla_jg_{(2)ki})\}e^{-2r}+..
\lb{C21}
\ee
As we see from (\ref{R2}) and (\ref{gdg}) the ($i,r$) component of the Ricci
tensor has expansion
\be
R_{ri}=[-\nabla_j g^j_{(2)i}+\partial_j \Tr g_{(2)}]e^{-2r}+..
\lb{R12}
\ee
Looking now at the expansion of $(ir)$ component of the Einstein equations
(\ref{3.5}) we get the constraint on the coefficient $g_{(2)ij}$
which can be presented in the form
\be
&&\nabla_j t^j_i=-{\beta\over 4}\epsilon_i^{\ j}\partial_j R~~, \nonumber \\
&&t_{ij}=g_{(2)ij}-g_{(0)}\Tr g_{(2)}+{\beta\over 2}(\epsilon_i^{\ k}g_{(2)jk}+\epsilon_j^{\ k}g_{(2)ik})~~.
\lb{t}
\ee
The holographic stress-energy tensor is defined as
\be
T_{ij}={1\over 8\pi G_N} t_{ij}~~.
\lb{T}
\ee
The part in the holographic stress tensor which is due to the Chern-Simons term is in agreement with
the general expression (\ref{ht5}).
The divergence of (\ref{T}) produces a gravitational anomaly
\be
\nabla_j T^j_i=-{\beta\over 32\pi G_N} \epsilon_i^{\ j}\partial_j R~~.
\lb{T1}
\ee
This is precisely the anomaly that is expected to appear in two dimensions.
It originates from $p_1$ (see \cite{Alvarez-Gaume:1983ig}, \cite{Leutwyler:1984nd})  via the descent mechanism outlined in section 2.

\bigskip

\noindent {\it Dimension d=6.} The analysis in six-dimensional case is much more
laborious. In the absence of the gravitational Chern-Simons term the analysis
was done in \cite{CMP}. The construction of the holographic stress tensor in
terms of the coefficients in the expansion of metric is then already
non-trivial and an explicit prescription is given in \cite{CMP}.
Turning on the Chern-Simons term makes things even more complicated. Fortunately for us
we do not need to go to the full analysis of the modified Einstein equations
but have to look only at the $e^{-6r}$ order of the $(ir)$ component of the
Einstein equations which determines, as was shown in \cite{CMP},  the conservation law for the holographic
stress-energy tensor.

The tensor $S^{\mu\nu\alpha}$ in six dimensions takes the form
\be
S^{\mu\nu\alpha}=-{1\over 2}\epsilon^{\mu_1..\mu_6 \mu}R^\nu_{\
  a_1\mu_1\mu_2}R^{a_1}_{\ a_2\mu_3\mu_4}R^{a_2\alpha}_{\ \ \mu_5\mu_6}~~.
\lb{S6}
\ee
The expansion (\ref{R1}), (\ref{R2}) and (\ref{R3}) of the Riemann tensor
is sufficient for the analysis of the leading behavior of the components of
the tensor (\ref{S6}).  Below we summarize this analysis:
\be
S^{rij}=\{-{1\over 2}\epsilon^{k_1..k_6}R^i_{\ n_1 k_1 k_2}R^{n_1}_{\ n_2k_3
  k_4}R^{n_2j}_{\ \ k_5k_6}-\epsilon^{k_1..k_5j}R^i_{\ n_1k_1k_2}R^{n_1}_{\
  n_2k_3k_4}g^{n_2}_{(2)k_5}\}e^{-8r}~~,
\lb{Srij}
\ee
\be
&&S^{irj}=\{\epsilon^{k_1..k_5i}g^{(2)}_{n_1k_1}R^{n_1}_{\ n_2k_2k_3}R^{n_2j}_{\
  \ k_4k_5}+2\epsilon^{k_1..k_4ji}g^{(2)}_{n_1k_1}R^{n_1n_2}_{\ \
  k_2k_3}g^{(2)}_{n_2k_4}\nonumber \\
&&+4\epsilon^{k_1..k_4ji}\nabla_{k_1}g^{(2)}_{nk_2}\nabla_{k_4}g^n_{(2)k_3}\}e^{-8r}~~,
\lb{Sirj}
\ee
\be
S^{rir}=\epsilon^{k_1..k_6}R^i_{\ n_1k_1k_2}R^{n_1}_{\ n_2 k_3
  k_4}\nabla_{k_5}g^{n_2}_{(2)k_6}\ e^{-8r}~~,
\lb{Srir}
\ee
\be
S^{kin}=\epsilon^{k_1..k_5k}R^i_{\ n_1k_1k_2}R^{n_1}_{\
  n_2k_3k_4}\nabla^ng^{n_2}_{(2)k_5} \ e^{-10r}~~,
\lb{Skin}
\ee
where we keep only the leading terms, components of the Riemann tensor are
defined with respect to metric $g_{(0)ij}$.
Notice that in (\ref{Skin}) we have dropped the terms that vanish when the trace
$g^{(0)}_{kn}S^{kin}$ is taken. Such terms appear both in the order $e^{-8r}$
and in the order $e^{-10r}$ and are not shown in (\ref{Skin}).

The expansion (\ref{Srij}), (\ref{Sirj}), (\ref{Srir}) and (\ref{Skin}) should
be now substituted into equation (\ref{C}). After some reshuffle and noticing
that quite a few terms vanish due to the Bianchi identities we get a quite
simple result
\be
&&C_{ri}=\nabla_j\{-{1\over 4} \epsilon^{k_1..k_6}R^i_{\ n_1k_1k_2}R^{n_1}_{\
  n_2k_3k_4}R^{n_2j}_{\ \ k_5k_6}\nonumber \\
&&+ {1\over 2}\epsilon^{k_1..k_5i}R^j_{\
  n_1k_1k_2}R^{n_1}_{\ n_2k_3k_4} \ g^{n_2}_{(2)k_5}+{1\over 2}\epsilon^{k_1..k_5j}R^i_{\
  n_1k_1k_2}R^{n_1}_{\ n_2k_3k_4} \ g^{n_2}_{(2)k_5}\} e^{-6r}~~.
\lb{C2}
\ee
Obviously, the first term in the brackets is antisymmetric in indices $i$ and
$j$ while the two other terms form a symmetric tensor. The latter will modify the
holographic stress-energy tensor while the first term will produce a
gravitational anomaly.

The expansion of the Ricci tensor to the required order was found in
\cite{CMP}. We refer the reader to that paper for the details. The result is
\be
R_{ri}=-3\nabla_j(g_{(6)}-A_{(6)}+{1\over 24}S)^j_{\ i} e^{-6r}~~,
\lb{Rir}
\ee
where we focus only on the term of the order $e^{-6r}$. The tensors
$A_{(6)ij}$ and $S_{ij}$ are local covariant functions of the metric $g_{(0)ij}$
and its derivative, exact expressions are rather lengthy and are given in paper \cite{CMP}.

Introduce tensor $t^{(\beta)}_{ij}$ as follows
\be
t^{(\beta)}_{ij}=g_{(6)ij}-A_{(6)ij}+{1\over 24}S_{ij}
-{\beta\over 3}\epsilon^{k_1..k_5}{}_{(i}
R_{j)  n_1k_1k_2}R^{n_1}{}_{ n_2k_3k_4} \ g^{n_2}_{(2)}{}_{k_5}~~,
\lb{tij}
\ee
where  the normalization has been chosen in agreement with \cite{CMP}.
Note that in  dimension $d>2$ the coefficient $g_{(2)}$ is a local covariant function of
the metric $g_{(0)}$. In particular, for $d=6$, we have that
\be
g_{(2)ij}=-{1\over 4}(R_{ij}-{1\over 10}Rg_{(0)ij})~~.
\ee
This relation  remains the same when the
Chern-Simons term is added to the bulk equations.
The constraint that comes from the $(ir)$-component of the Einstein equation, $R_{ir}+\beta C_{ir}=0$, can
be now presented in the following form
\be
\nabla_jt^j_{(\beta)i}=-{\beta\over 12}\epsilon^{k_1..k_6}\nabla_j(R^i_{\ n_1k_1k_2}R^{n_1}_{\
  n_2k_3k_4}R^{n_2j}_{\ \ k_5k_6})~~.
\lb{tt}
\ee
Obviously, the tensor $t^j_{(\beta)i}$ (\ref{tij}) is defined by this equation
only up to a covariantly conserved term
(proportional to $h_{(6)}$).
The holographic stress-energy tensor in the absence of the Chern-Simons term, $\beta=0$ in this case,
was defined in \cite{CMP}. Extending this definition to the present case and taking into account a general
expression (\ref{ht5}) for the CS contribution we define the stress-energy tensor
as follows
\be
T_{ij}={3\over 8\pi G_N}t^{(\beta)}_{ij}~~.
\lb{TT}
\ee
Defined this way this tensor (in the case when $\beta=0$) was shown in \cite{CMP} to be symmetric, covariantly conserved and
its trace to be the conformal anomaly of the boundary CFT.
Notice, that the
$\beta$-dependent modification in (\ref{tij}) is traceless so that the trace
of the modified stress-energy tensor remains the same.  The stress tensor is
however not conserved anymore due to the gravitational anomaly,
\be
\nabla_jT^j_i=-{\beta\over 32\pi G_N} \epsilon^{k_1..k_6}\nabla_j(R^i_{\ n_1k_1k_2}R^{n_1}_{\
  n_2k_3k_4}R^{n_2j}_{\ \ k_5k_6})~~.
\lb{TTT}
\ee
This is exactly the anomaly that originates in the descent mechanism from the term $\Tr R^4$ in
the Pontryagin class $p_2$. This is however not the most general form of the gravitational anomaly
in six dimensions. Indeed, another possible  anomaly  originates from the term $(\tr R^2)^2$ that appears both
in $p_2$ and in $p_1^2$. This anomaly comes out holographically if one adds a
reducible form of Chern-Simons term
to the bulk gravitational action.

\bigskip

\noindent {\bf Anomaly from the reducible Chern-Simons term.} In six dimensions the only possible reducible form of the
Chern-Simons term is $W^{(1,1)}_{\tt CS}=4a_3\int  \Omega_3 \wedge d\Omega_3$. Adding this term to the gravitational action
\be
W_{\tt gr}=W_{\tt EH}-{\beta\over 32\pi G_N}W_{\tt CS}-{\beta_1\over 128\pi G_N}W^{(1,1)}_{\tt CS}
\lb{action}
\ee
with some coupling $\beta_1$ we get, after some regrouping the terms, the modified Einstein equations
\be
R_{\mu\nu}=-d G_{\mu\nu}-\beta C_{\mu\nu}-\beta_1 C^{(1,1)}_{\mu\nu}~~,
\lb{ME}
\ee
where we took into account that $C^{(1,1)}_{\mu\nu}$ is traceless.
Tensor $C^{(1,1)}_{\mu\nu}$ was defined in (\ref{2.13}) to take the form
\be
C^{\mu\nu}_{(1,1)}=-{1\over 2}\epsilon^{\sigma_1..\sigma_6(\mu}\nabla_\alpha [R^{\nu)\alpha}_{\ \ \sigma_1 \sigma_2}
(R^{c_1}_{\ c_2\sigma_3\sigma_4} R^{c_2}_{\ c_1\sigma_5\sigma_6})]~~,
\lb{C11}
\ee
where all indices run from 1 to 7.
Again we have to look at the $(r,i)$ component of the modified Einstein equations (\ref{ME}). The analysis
goes through the same steps as before, now for the tensor $C_{(1,1)}^{\mu\nu}$. Skipping the details which are
pretty straightforward we present the result for the leading term in the large $r$ expansion
\be
C^{ri}_{(1,1)}=\nabla_j \{[-{1\over 4}\epsilon^{k_1..k_6}R^{ij}_{\ \ k_1 k_2}+\epsilon^{k_2..k_6(i}g^{j)}_{(2)k_2}]
R^{n_1}_{\ n_2 k_3k_4}R^{n_2}_{\ n_1 k_5 k_6} \} e^{-8r}~~.
\lb{C1,1}
\ee
Here all indices (including $n_1$ and $n_2$) run from 1 to 6.
The new constraint which comes from the $(r,i)$ component of  equations (\ref{ME})
can be properly formulated in terms of the tensor
\be
t_{(\beta,\beta_1)}^{ij}=t_{(\beta)}^{ij}-{\beta_1\over 3}\epsilon^{k_2..k_6(i}g^{j)}_{(2)k_2}
R^{n_1}_{\ n_2 k_3k_4}R^{n_2}_{\ n_1 k_5 k_6}~~,
\lb{tb}
\ee
where in the last term the symmetrization in indices $i$ and $j$ is assumed.
We can now define the holographic stress tensor as
\be
T_{ij}={3\over 8\pi G_N}t^{(\beta,\beta_1)}_{ij}~~
\lb{Tb}
\ee
in analogy with (\ref{TT}).
Its divergence is now a combination of the contributions from both the reducible and irreducible Chern-Simons terms
\be
\nabla_j T^j_{\ i}=-{\beta\over 32\pi G_N}\epsilon^{k_1..k_6}\nabla_j(R^i_{\ n_1k_1k_2}R^{n_1}_{\
  n_2k_3k_4}R^{n_2j}_{\ \ k_5k_6})\nonumber \\
-{\beta_1\over 32\pi G_N}\epsilon^{k_1..k_6}\nabla_j(R^{ij}_{\ \ k_1 k_2}
R^{n_1}_{\ n_2 k_3k_4}R^{n_2}_{\ n_1 k_5 k_6})~~.
\lb{Anomaly}
\ee
The second term in the right hand side of (\ref{Anomaly}) is precisely a contribution to the gravitational anomaly
from the term $(\Tr R^2)^2$ via the descent method.
So that equation (\ref{Anomaly}) presents the most general
form of the gravitational anomaly in six dimensions.

\bigskip

\noindent {\bf Some comments.} The tensor (\ref{TT}), or more generally (\ref{Tb}),
has a dual meaning. It is the expectation value of the quantum stress-energy
tensor in the dual CFT and is the quasi-local stress-energy tensor
introduced by York and Brown \cite{BY} to define the energy and angular
momentum for a
solution to the bulk gravitational equations. In three dimensions the Cotton
tensor vanishes for any metric conformal to the constant curvature metric.
That's why the BTZ metric describing a three-dimensional
black hole remains a solution to the modified Einstein equations (\ref{3.5}).
The stress-energy tensor (\ref{T}), (\ref{t}) then can be used to calculate
the modified values for the mass and angular momentum of the BTZ black hole
\cite{Kraus}, \cite{SS}. In higher dimensions a general solution to the Einstein
equations with a cosmological term is no more maximally symmetric metric
so that the tensor $C_{\mu\nu}$ is non-vanishing. This means that some modification
of the known solutions describing a black hole in asymptotically AdS space-time
should be expected. The finding exact solutions to the modified Einstein
equations (\ref{3.5}) or (\ref{ME}) is an interesting problem that possibly can
be approached numerically. Provided such a solution is known our formulas (\ref{T}),
(\ref{t}) or (\ref{Tb}), (\ref{tb}) can be used to calculate the conserved
quantities of the solution. We however note that unlike the three-dimensional
case in higher dimensions the $\beta$-dependent
modification in (\ref{tb}) vanishes if the boundary metric $g^{(0)}_{ij}(x)$
is flat or is a maximally symmetric constant curvature  metric.
Only if there is a solution  which approaches
a non-maximally symmetric metric at infinity then the modification (\ref{tb})
or (\ref{tij}) of the
stress-energy tensor would be relevant. Also, only in this case the
gravitational anomaly (\ref{Anomaly}) will be actually visible.

\section{Remarks on anomalies}

\medskip

\noindent {\bf Comparison with the descent method.} The Chern-Simons term that was added to the bulk
gravitational action can be used
for calculation of the anomaly using the descent method. In this subsection we do this calculation and compare the
resultant anomaly with the one obtained holographically and find that these two anomalies are identical.
It is more convenient to calculate first the local Lorentz anomaly and then transform the result to the gravitational anomaly.

We start with some general remarks on the Lorentz symmetry and the Lorentz anomaly.
We introduce the vielbein stress-energy action as $T^{i}_{(h)a}={2\over h}{\delta W\over \delta h^a_i}$.
The subscript $(h)$ is supposed to differ this from the metric stress tensor
$T^{ij}_{(g)}={2\over \sqrt{g}}{\delta W\over \delta g^{ij}}$. $W$ is the action of the theory in question. These two objects
are related as
$$
T_{(h)}^{ia}=T^{ij}_{(g)}h^a_j+T^{ji}_{(g)}h^a_j~~.
$$
We raise the Lorentz indices with the help of $\delta^{ab}$. Under the infinitesimal local Lorentz transformations
the vielbein and the spin connection transform as
\be
\delta h^a_i=\alpha^{a}_{\ b}h^b_i~,~~\delta \omega^a_{\ b,i}=-\partial_i \alpha^a_{\ b}~, ~~\alpha_{ab}=-\alpha_{ba}~~.
\lb{LL}
\ee
In the Lorentz invariant theory one has that $T_{(h)}^{[ab]}=0$, $T^{ab}_{(h)}=T^{ia}_{(h)}h^b_i$.
In d-dimensional quantum chiral theory the Lorentz symmetry may be violated if $d=4k-2$.
In the descent method the violation is determined by a $(d+2)$-dimensional invariant form $I_{d+2}$ that is  polynomial
in the Riemann curvature as was explained in section 2. This form is locally exact $I_{d+2}=dI_{\tt CS}$,
where $I^{(d+1)}_{\tt CS}$ is a $(d+1)$-dimensional Chern-Simons term. Under the local Lorentz transformations (\ref{LL}) this term
changes as $\delta_\alpha I^{(d+1)}_{\tt CS}=d[X^{ab}\alpha_{ba}]$, where $X^{ab}= X^{ab}_{i_1..i_d}dx^{i_1}\wedge
..\wedge dx^{i_d}$ is a $d$-form. The anomaly then shows up in the non-vanishing  antisymmetric part
of $T^{ab}_{(h)}$ and reads
\be
{1\over 2}T^{[ab]}_{(h)}=\epsilon^{i_1..i_d}X^{ab}_{i_1..i_d}~~.
\lb{Tab}
\ee
For instance, take  $I^{(d)}_{\tt CS}=-{\beta\over 32\pi G_N}a_n\Omega^{(2n+1)}_{\tt CS}$ and apply the descent procedure.
Using (\ref{2.8}) we get that it leads to the Lorentz anomaly
\be
{1\over 2}T^{[ab]}_{(h)}=-{\beta\over 32\pi G_N}\epsilon^{i_1..i_d}(R^{a}_{\ c_1,i_1i_2}..R^{bc_{d-2}}_{\ \ i_{d-1}i_d})~~,
\lb{TL}
\ee
the expression in the right hand side is obviously antisymmetric in indices $a$ and $b$ if $d=4k-2$.
The non-vanishing antisymmetric part of $T^{ab}_{(h)}$ would imply that
the metric stress-energy tensor is not symmetric. Keeping $T^{ij}_{(g)}$ symmetric we should subtract
the antisymmetric part ${1\over 2}h^i_ah^j_bT^{[ab]}_{(h)}$. The resultant
symmetric tensor is not conserved,
\be
\nabla_jT^{ij}_{(g)}=-{\beta\over 32\pi G_N}\epsilon^{i_1..i_d}\nabla_j(R^{i}_{\ c_1,i_1i_2}..R^{jc_{d-2}}_{\ \ i_{d-1}i_d})~~.
\lb{TTT1}
\ee
This result is the same as if we replaced $R_{ab}\rightarrow R_{ab}+2\nabla_{[a}\xi_{b]}$ in $I_{d+2}$ and looked at the
first order in $\xi_a$ term. This latter prescription was given in section 2.
Comparing (\ref{TTT1}) to the holographic expressions (\ref{T1}) ($d=2$) and (\ref{TTT}) ($d=6$) we see that
in two different methods,  by adding the Chern-Simons action $\int I^{(d+1)}_{\tt CS}$
to the bulk gravitational action and looking at the divergence of the dual
stress tensor in the holographic method  and, in the second method, by using the same form $I^{(d+1)}_{\tt CS}$ in the descent procedure,
we get same result. Same is true for the anomaly determined by the reducible form
(verified for $\Omega_3d\Omega_3$ when $d=7$) of the Chern-Simons action. Here we have checked this by  brute force.
However, it seems that there may be a more general proof that two methods lead to
same result\footnote{I thank Jan de Boer and Kostas Skenderis for suggesting this to me.}.
It would be interesting to understand this issue.

\bigskip

\noindent {\bf Anomaly in $(2,0)$ six-dimensional conformal theories.} In six
dimensions there are two known
$(2,0)$ supersymmetric conformal theories. The first one is the free tensor
multiplet theory which describes the low energy dynamics of a single M5 brane.
The other one is the strongly interacting $(2,0)$ conformal theory
describing $N$ coincident M5 branes. Some information about this second
theory can be gained from its conjectured holographic duality to M-theory (or, in large $N$,
limit to the 11-dimensional supergravity) on $AdS_7\times  S^4$ background.
In particular, the holographic  anomalies  is an important source of
information about the theory.
The conformal anomaly in the $(2,0)$ theory was calculated in
\cite{Henningson:1998gx}. The comparison to the
anomaly in the free tensor multiplet was done for instance in
\cite{Bastianelli:2000hi}. The conformal anomaly in two theories are  mainly related
by factor $4N^3$. This is the leading contribution to the anomaly which
holographically originates from the tree level supergravity action linear in
the curvature. The one-loop effective action contains quartic in
curvature terms. They  lead to the $O(N)$ modification of the anomaly.
A nice discussion of this can be found in \cite{Tseytlin:2000sf}.

The maximal $(2,0)$ supersymmetric theories are necessarily chiral
so that the gravitational anomaly is expected to appear. The free tensor
multiplet consists of 5 scalars, a (anti)selfdual antisymmetric tensor and 2
Weyl fermions.  The gravitational anomaly is thus a descent of
8-form
\be
I_8^{\tt tens}=I_A+2I_{1/2}=-{i\over (2\pi )^3 192}[\Tr R^4-{1\over
  4}(\Tr R^2)^2]~~,
\lb{I8t}
\ee
where we use  formulas of section 2. The corresponding anomaly of interacting $(2,0)$,
as conjectured in \cite{Harvey:1998bx} (by assuming that the  M5-brane
anomaly should be compensated by the inflow anomaly),  is determined by
\be
I_8^{(2,0)}=N I_8^{\tt tens}~~.
\lb{I820}
\ee
Note that we focus on the gravitational part of the anomaly neglecting the
gauge field and the mixed anomalies.

The anomaly (\ref{I820}) is subleading in $N$ that means that holographically
it originates from a one-loop term in the effective action.
Terms of this type were studied in \cite{Tseytlin:2000sf}. There are few
terms in the one-loop action which are quartic in curvature. The one of our interest
contains invariant $\Tr R^4-{1\over 4}(\Tr R^2)^2$ that is exactly of the
type that appears in (\ref{I820}), (\ref{I8t}). More precisely one finds
(we use notations of \cite{Tseytlin:2000sf} and make the continuation to Euclidean signature)
\be
W=-{i\over (2\pi)^4 \cdot 3\cdot 2^6} T_2\int {\cal C}_3\wedge (\Tr R^4-{1\over 4}(\Tr R^2)^2)~~,
\lb{WM}
\ee
where $T_2$ is the membrane tension and ${\cal C}_3$ is 3-form potential, for the 11-dimensional action. This term
was first derived in \cite{Duff:1995wd} and plays an important role in the inflow mechanism
\cite{Witten:1996hc}, \cite{Freed:1998tg}.
 Here we follow the line of
reasoning suggested in \cite{Kraus:2005vz}. We first integrate (\ref{WM}) by
parts and then  compactify on $S^4$ with flux\footnote{This is the right period for the form $F$, as was discussed in
\cite{Duff:1995wd}.}
$T_2 \int_{S^4}F=2\pi N$, $F=d{\cal C}_3$. The term (\ref{WM}) then reproduces exactly the Chern-Simons action
 to be  added to the 7-dimensional  gravitational action.
This action is a source of the six-dimensional gravitational anomaly either through the holographic
procedure or in the descent method analysis. The anomaly takes exactly the form conjectured in
 \cite{Harvey:1998bx}.
We can now determine explicitly values of the couplings $\beta$ and
$\beta_1$ in the Chern-Simons action.
In the units in which radius of $S^4$ is $1/2$ we have that
$16\pi G^{(7)}_N={3\pi^3/N^3}$ (see \cite{Tseytlin:2000sf}) and hence $\beta={i\over 2^9 N^2}$ and $\beta_1=-{i\over 2^{11}N^2}$.

\bigskip

\bigskip

\noindent {\large \bf Acknowledgments}

\bigskip

\noindent
The author is grateful to M. Anderson, G. Arutyunov, J. de Boer, R. Gover,
R. Graham, R. Helling,
Yu. Obukhov and K. Skenderis for helpful discussions and correspondence.
This work is supported in part by  DFG grant Schu 1250/3-1.

\appendix{Curvature components}
\setcounter{equation}0

For metric (\ref{3.6}) the components of the $(d+1)$-dimensional  Riemann tensor are
\be
R^r_{\ irj}&=&{1\over 2}[-g''+{1\over 2}g'g^{-1}g']_{ij} \nonumber \\
R^r_{\ ikj}&=&-{1\over 2}[\nabla_kg'_{ij}-\nabla_jg'_{ik}] \nonumber \\
R^{l}_{\ ikj}&=&R^{l}_{\ ikj}(g)-{1\over 4}g'_{ij}g^{ln}g'_{nk}+{1\over 4}g'_{ik}g^{ln}g'_{nj}~~,
\lb{a1}
\ee
where $g'\equiv \partial_r g$.
Components of Ricci tensor are
\be
R_{ij}&=&R_{ij}(g)-{1\over 2} g''_{ij}-{1\over 4}g'_{ij}\Tr(g^{-1}g')+{1\over 2}(g'g^{-1}g')_{ij} \nonumber \\
R_{ri}&=&{1\over 2}[\nabla_k(g^{-1}g')^k_{i}-\nabla_i\Tr(g^{-1}g')] \nonumber \\
R_{rr}&=&-{1\over 2}\Tr(g^{-1}g'')+{1\over 4}\Tr(g^{-1}g'g^{-1}g')
\lb{a2}
\ee
and the Ricci scalar is
\be
R=R(g)-\Tr(g^{-1}g'')-{1\over 4}[\Tr(g^{-1}g')]^2+{3\over 4}\Tr(g^{-1}g'g^{-1}g')~~.
\lb{a3}
\ee

\bigskip

\appendix{Expansion for the inverse metric  and the Riemann tensor}
\setcounter{equation}0

As preparation we present here expressions for the inverse of effective metric $g_{ij}(r,x)$
and  its derivatives with respect to $r$
\be
&&g^{-1}=e^{-2r}g^{-1}_{(0)}[1-g_{(2)}e^{-2r}+ (-g_{(4)}+g_{(2)}g^{-1}_{(0)} g_{(2)})e^{-4r} \nonumber \\
&&+(-g_{(6)}+ g_{(2)}g^{-1}_{(0)} g_{(4)}   +  g_{(4)}g^{-1}_{(0)} g_{(2)} -g_{(2)}g^{-1}_{(0)}g_{(2)}g^{-1}_{(0)} g_{(2)})e^{-6r}+..]
       g^{-1}_{(0)} \nonumber \\
&&g'=2e^{2r}(g_{(0)}-g_{(4)}e^{-4r}-2g_{(6)}e^{-6r}+..) \nonumber \\
&&g''=4e^{2r}(g_{(0)}+g_{(4)}e^{-4r}+4g_{(6)}e^{-6r}+..)~~,
\lb{gg}
\ee
where $..$ stands for the sub-leading terms.
In particular we have that
\be
&&g^{-1}g'=2 g^{-1}_{(0)}\{ 1-g_{(2)}e^{-2r}+(-2g_{(4)}+g_{(2)}g^{-1}_{(0)} g_{(2)})e^{-4r}\nonumber \\
&&+((-3g_{(6)}+ 2g_{(2)}g^{-1}_{(0)} g_{(4)}   +  g_{(4)}g^{-1}_{(0)} g_{(2)} -g_{(2)}g^{-1}_{(0)}g_{(2)}g^{-1}_{(0)} g_{(2)})e^{-6r}+..\}
\lb{gdg}~~.
\ee
It is important to note that if the dimensions $d$ is even there generally appears a logarithmic\footnote{The logarithm appears if one
uses the radial coordinate $\rho=e^{-2r}$ instead of $r$.}
term $h_{(d)}re^{-(d-2)r}$  in the expansion (\ref{3.7}). In (\ref{gdg}) this would add extra terms
$g^{-1}_{(0)}(h_{(d)} e^{-dr}-(d-2)h_{(d)}re^{-dr})$ plus the corresponding higher order terms.

Using  (\ref{a1}) we  get an expansion for the  components of the Riemann tensor,
\be
&&R^{r}_{\ irj}=-(g_{(0)}e^{2r}+g_{(2)})_{ij}+..\\
&&R^{ri}_{\ \ rj}=-\delta^i_j+O(e^{-4r})  \nonumber
\lb{R1}
\ee
\be
&&R^r_{\ ikj}=(\nabla_kg^{(2)ij}-\nabla_jg^{(2)ik})+.. \\
&&R^{ri}_{\ \ kj}=(\nabla_kg_{(2)j}^{i}-\nabla_jg_{(2)k}^{i})e^{-2r}+.. \nonumber \\
&&R_{kj}^{\ \ ri}=(\nabla^kg^{(2)i}_{j}-\nabla^jg^{(2)i}_{k})e^{-4r}+.. \nonumber
\lb{R2}
\ee
\be
R^l_{\ ikj}&=&(g_{(0)ik}\delta^l_j-g_{(0)ij}\delta^l_k)e^{2r}+(R^{(0)l}_{\ \ ikj}+g_{(0)ij}g_{(2)k}^{l}-
g_{(0)ik}g_{(2)j}^l)+.. \\
R^{li}_{\ \ kj}&=&(\delta^i_k\delta^l_j-\delta^i_j\delta^l_k)+(R^{(0)li}_{\ \ \ kj}+\delta^i_jg_{(2)k}^{l}-
\delta^i_kg_{(2)j}^l -\delta^l_jg_{(2)k}^{i}+
\delta^l_kg_{(2)j}^i)e^{-2r}+.. \nonumber
\lb{R3}
\ee
We use the inverse metric $g^{ij}_{(0)}$ to raise the indices. An  expansion for the Levi-Civita symbol is
\be
\epsilon^{i_1..i_dr}=e^{-dr}\epsilon^{i_1..i_d}_{(0)}+..~~,
\lb{LC}
\ee
where $\epsilon^{i_1..i_d}_{(0)}$ is defined with respect to the metric $g_{(0)ij}$.

\end{document}